%% file: SpaceTimeDynamicsEstimation_accepted.tex
\documentclass{aa}

\usepackage{txfonts}
\usepackage{amssymb}
\usepackage{amsmath}
\usepackage{wasysym}
\usepackage{natbib}
\usepackage{subfigure}
\usepackage{graphicx}
\usepackage{wrapfig}
\usepackage{lscape}
\usepackage{rotating}
\usepackage{appendix}
\usepackage{epstopdf}
\usepackage{epsfig}

\begin{document}

   \title{Space-time Dynamics Estimation from Space Mission Tracking Data}

   \author{D. Dirkx
          \inst{1,2}
          \and
          R. Noomen 
          \inst{1}
          \and
         P.N.A.M. Visser
          \inst{1}
          \and
          L. Gurvits
          \inst{2,1} 
          \and
          L.L.A. Vermeersen
          \inst{1} 
          }

   \institute{Delft University of Technology,
               Kluyverweg 1, 2629HS Delft, The Netherlands\\
              \email{d.dirkx@tudelft.nl}
         \and
             Joint Institute for VLBI ERIC,
             PO Box 2, 7990 AA, Dwingeloo, The Netherlands
             }

%   \author{D, Dirkx
%          \inst{1,2}
%          \and
%          R. Noomen 
%          \inst{2}
%          }
%
%   \institute{Delft University of Technology,
%              Kluyverweg 1, 2629HS Delft, The Netherlands\\
%              \email{d.dirkx@tudelft.nl}
%         \and
%             Joint Institute for VLBI in Europe,
%            Kluyverweg 1, 2629HS Delft, The Netherlands\\
%             \email{d.dirkx@tudelft.nl}
%             }

   \date{Received 8 October 2015}

\abstract
{}
{Many physical parameters that can be estimated from space mission tracking data influence both the translational dynamics and proper time rates of observers. {These} different proper time rates cause a variability of the time transfer observable beyond that caused by their translational (and rotational) dynamics. With the near-future implementation of {(interplanetary) transponder} laser ranging, these effects will become increasingly important, requiring a re-evaluation of {the} common data analysis practice of using \textit{a priori} time ephemerides, which is the goal of this paper.}
{We  develop a framework for the simultaneous estimation of {the} initial translational state and {the} initial proper time of an observer, {with the goal of} facilitating robust tracking data analysis from next-generation space missions carrying highly accurate clocks and tracking equipment. {Using our approach, the influence of physical parameters on both translational and time dynamics are considered at the same level in the analysis, and mutual correlations between the signatures of the two are automatically identified}. We perform a covariance analysis using our proposed method with simulated laser data from Earth-based stations to both a Mars and Mercury lander.}
{{Using four years of tracking data for the Mars lander simulations, we find a difference between our results using the simultaneous space-time dynamics} estimation and the classical analysis technique (with an \textit{a priori} time ephemeris) of around 0.1 \% in formal errors and correlation coefficients. For a Mercury lander this rises to around 1\% for a 1-month mission and 10 \% for a 4-year mission.{ By means of Monte Carlo simulation, we find }that using an \textit{a priori} time ephemeris of representative accuracy  will result in estimation errors that are orders of magnitude above the formal error when processing highly accurate laser time transfer data.}
{}


\maketitle

% \linenumbers

%% main text
\input{mainText}

\bibliographystyle{aa} 
\bibliography{Bibliography/Bibliography}

% \bibitem[ ()]{}

\end{document}

%% file: mainText.tex
\section{Introduction}
\label{paper4sec:Introduction}

{By using highly stable space- and Earth-based clocks \citep[\textit{e.g.}, ][]{DelvaEtAl2012,BloomEtAl2014},} combined with {accurate} time transfer techniques, { \citep[\textit{e.g.}, ][]{Bjerhammar1985,Klioner1992,PetitWolf1994,BlanchetEtAl2001}, the behaviour of distant clocks can be accurately compared}. {Due to relativistic clock effects, the data that are obtained from such experiments can be used for improving both} geodetic measurements \citep{MullerEtAl2008} and experimental tests of general relativity \citep{Angelil2014}. {Key examples of such experiments in ground-to-space time transfer include the T2L2 project on the Jason-2 satellite \citep{ExertierEtAl2010} and the upcoming ACES project \citep{DelvaEtAl2012}.}

{Time transfer data analysis can potentially be complicated by the fact that this analysis} is contingent upon accurate knowledge of the state history of the transmitter and the receiver. It was found by \cite{DuchayneEtAl2009}, though, that the influence of orbital errors on the achievable accuracy of time transfer is acceptably small for the case of the ACES project. {Therefore, an \textit{a priori} orbital solution can be used for the robust analysis of the ACES clock data. Consequently, these data need not (and cannot) be used to improve the orbital solution of the space segment. This approach of decoupled time dynamics and translational dynamics (using an \textit{a priori} orbital solution for clock data analysis) is typical, and largely sufficient, in the analysis of data from current space missions. However, this is not necessarily true for all current and future time transfer experiments, including those seeking to use relativistic clock effects for improved science return.} For instance, it is found by \cite{DelvaEtAl2015} that orbital errors{ of the Galileo 5 and 6 satellites (in eccentric Earth orbits) }will manifest {themselves} in the analysis of the data from the clocks of these satellites.{ This indicates that the typical approach of using an \textit{a priori} orbital solution for the clock data analysis is not necessarily the optimal strategy for all present and future missions. }

{For cases where the nominal translational dynamics solution is insufficiently accurate to remove all observable relativistic clock signatures, these clock effects could also be used to improve the orbit determination and parameter estimation quality. However, this requires a suitable inclusion of the clock comparison data in the orbit determination process. Crucially, the clock behaviour needs to be estimated consistently and concurrently with the translational state behaviour.
}
When using the same data types for orbit determination and time transfer, {a concurrent analysis of orbital and time behaviour may be especially important. This will be the case for missions employing interplanetary laser ranging (ILR).} This technology could be used for range measurements at the 1 mm to 1 cm level \citep{Degnan2002,TuryshevEtAl2010,DirkxEtAl2014}{, and time transfer at the resulting 3-30 ps level (assuming sufficiently accurate hardware is used). }%, compared to a typical current and near-future radiometric range measurement accuracues of 1 m \citep{ThorntonBorder2000} and 20 cm \citep{IessEtAl2014}, respectively. 
{The need for very accurate clock models is especially high for missions employing a one-way (radio or optical) data type  \citep{ElyEtAl2014,DirkxEtAl2015}, as clock effects (both noise and relativistic) then accumulate over long time periods.}

{There are various proposals to exploit space-based clocks to test general relativity, using both near-Earth and interplanetary spacecraft \citep{Angelil2014,ScharerEtAl2014,DengXie2013a,PanXie2015}. Such tests could be performed using both optical and radiometric systems. An ILR system, however, will be able to provide the highest accuracy for interplanetary time transfer. The analysis of the performance of these missions requires an analysis of the influence of inaccuracies in the translational dynamics on the analysis of the clock data, as was done for the ACES project by \cite{DuchayneEtAl2009}. Conversely, it is necessary to analyze the degree to which the clock signal contributes to the estimation of the physical parameters of interest, compared to the measurable influence on the translational dynamics \citep{DengXie2013b,DengXie2013a}.}

Here, we propose a novel general analysis framework in which the estimation of both translational dynamics and proper time dynamics is simultaneously performed, along with the estimation of physical parameters of interest. To achieve this, we extend the typical orbit determination procedure \citep[\textit{e.g.,}][]{MontenbruckGill2000}{ to include the estimation of the initial proper times of the observers. }

{This approach is distinct from the current reduction of space mission tracking data, in which the relativistic clock corrections are applied in an \textit{a priori} manner (or updated iteratively during the orbit determination). Although this approach has been largely sufficient for past and present space missions, future missions (such as those employing ILR) may require a more detailed analysis of the combined influence of relativistic clock effects and translational dynamics on the tracking data. }

{Our proposed concurrent estimation will allow for a more robust analysis of ranging and time transfer data,} preventing signal from relativistic clock effects to be {unduly} interpreted as translational dynamics or \textit{vice versa}. Similarly, it will allow a direct quantification of the correlation between the signal of translational and time dynamics. {As has been shown by \cite[\textit{e.g.}, ][]{Iorio2012, VermaEtAl2014, FiengaEtAl2015, DengXie2015}, a robust quantification of the correlation between the parameters of interest is key in interpreting the estimation results in solar system tests of relativity.}

We apply our framework to two interplanetary missions: the tracking of a lander on Mars and on Mercury. The goal of our simulations is to demonstrate the application of our proposed methodology. {The goals is not} to provide a detailed analysis of the scientific performance of such mission, as is done by \textit{e.g.}, \cite{HeesEtAl2012}.

We start by reviewing the relevant details of relativistic celestial mechanics, time dynamics and  the fsormulation of the time-transfer observable in Section \ref{paper4sec:dynObs}. Our new derivation of the extended variational equations and the partial derivatives of the time-transfer observable, including the coupling between translational and time dynamics, is given in Section \ref{paper4sec:coupledEstimation}. The results of this section comprise the key novel theoretical aspects of this paper. We demonstrate our methodology with numerical simulations of both a Mars and a Mercury lander in Section \ref{paper4sec:testCases}, followed by our overall conclusions in Section \ref{paper4sec:conclusions}. 

\section{Dynamics and observable modelling}
\label{paper4sec:dynObs}
In this section,{ we provide an overview of the relevant models that we use in the development our concurrent estimation in Section \ref{paper4sec:coupledEstimation} and our simulations in Section \ref{paper4sec:testCases}. We} define our notation and conventions in Section \ref{sec:notConv}. We review the required models for the translational dynamics, time dynamics and range/time transfer observable in Sections \ref{paper4sec:relCelMech}, \ref{paper4sec:properTimeDynamics} and \ref{paper4sec:timeTransferObservable}, respectively.  

\subsection{Notation and conventions}
\label{sec:notConv}
We will use the ($-+++$) metric signature {and} the Einstein summation convention{. We} use Latin indices $i,j,k={1,2,3}$ to denote space indices and Greek indices $\mu,\nu,\sigma={0,1,2,3}$ to denote space-time indices (with 0 the time index). A subscript $_{,\mu}$ will denote partial differentiation w.r.t. $x^{\mu}$. 

We will {use} the post-Newtonian assumption of slow motion ($v\ll c$) and weak fields ($|g_{\mu\nu}-\eta_{\mu\nu}|\ll 1$), since our application of orbit determination is limited to solar system situations. As a result, the metric $g_{\mu\nu}$ can be written in terms of the metric perturbation $h_{\mu\nu}$ as follows \citep{KopeikinEtAl2011}:
\begin{equation}
g_{\mu\nu}=\eta_{\mu\nu}+h_{\mu\nu}\label{eq:paper4metricPerturbation}
\end{equation}
with $\eta_{\mu\nu}$ the Minkowski metric. We perform our analysis in the Barycentric Celestial Reference System (BCRS), which is kinematically non-rotating \citep{KopeikinVlasov2004}.

{We use the metric for} the solar system {at} first post-Newtonian order {that} is recommended by the IAU and discussed by \cite{SoffelEtAl2003}. However, we stress that our overall approach laid out in Section \ref{paper4sec:coupledEstimation} is not limited to this particular choice of metric, or even to metric theories of gravity. 

\subsection{Relativistic celestial mechanics}
\label{paper4sec:relCelMech}
In this section we will review the {aspects of relativistic }celestial mechanics necessary for our subsequent discussion. Further extensive discussion on the topic can be found in \textit{e.g.},\cite{Brumberg1972,Soffel1989,Brumberg1991,KopeikinVlasov2004,KopeikinEtAl2011}. 

The general four-dimensional equation of motion of a point mass {experiencing only gravitational forces} is \citep{KopeikinEtAl2011}:
\begin{align}
\frac{d^{2} x^{\mu}}{d \tau^{2}}&=-\Gamma^{\mu}_{\alpha\beta}\frac{d x^{\alpha}}{d \tau}\frac{d x^{\beta}}{d \tau}\label{eq:paper4fundamentalEomIncludingNonGrav}\\
\Gamma^{\mu}_{\alpha\beta}&=\frac{1}{2}g^{\mu\nu}\left(g_{\nu\alpha,\beta}+g_{\nu\beta,\alpha}-g_{\alpha\beta,\nu}\right)\label{eq:paper4christoffelMetricLink}
\end{align}
where $x^{\mu}$ denotes the four-position $\{ct,x,y,z\}$, $\tau$ denotes the proper time of an observer co-moving with the test body and $\Gamma^{\mu}_{\alpha\beta}$ denotes the Christoffel symbols of the metric connection. We will apply this equation in this article, noting that the additional terms that arise due to non-gravitational forces or self-gravitation \citep{KopeikinVlasov2004} may be included when requiring a {more} detailed dynamical model.

We also note that in certain alternative theories of gravity, the formulation of Eqs. (\ref{eq:paper4fundamentalEomIncludingNonGrav}) and (\ref{eq:paper4christoffelMetricLink}) {are} different \citep{KopeikinEtAl2011}, and additional terms need to be included. The inclusion of such terms will result in a modification of the equations that we use and derive further in this article. However, it will not change the general methodology that is derived in Section \ref{paper4sec:coupledEstimation}.

The equations of motion given by Eq. (\ref{eq:paper4fundamentalEomIncludingNonGrav}) may be written in terms of a coordinate time $t$ (such as Barycentric Coordinate Time TCB) instead of proper time $\tau$. This yields equations of motion expressed in a {timescale} typically used in orbit propagation and determination \textit{e.g.}, \citep{KopeikinEtAl2011}:
\begin{align}
\ddot{x}^{i}&=-\Gamma^{i}_{\alpha\beta}\dot{x}^{\alpha}\dot{x}^{\beta}+\frac{1}{c} \Gamma_{\alpha\beta}^{0}\dot{x}^{\alpha}\dot{x}^{\beta}\dot{x}^{i}\label{eq:paper4condensedEom}\\
&=-c^{2}\Gamma^{i}_{00}-2c\Gamma^{i}_{0j}\dot{x}_{j}-\Gamma^{i}_{jk}\dot{x}^{j}\dot{x}^{k}+\dot{x}^{i}\cdot...\nonumber\\
&\hspace{2cm}...\cdot\left(c\Gamma^{0}_{00}+2\Gamma^{0}_{0j}\dot{x}^{j}+\frac{1}{c}\Gamma^{0}_{jk}\dot{x}^{j}\dot{x}^{k}\right)\label{eq:paper4expandedEom}
\end{align}
where the overdot denotes a derivative w.r.t. coordinate time $t$. 

In numerical ephemerides of solar system bodies, the independent variable is typically Dynamical Barycentric Time (TDB), which is a scaled version of TCB and not a coordinate time itself.
In the remainder of this paper, we will assume that the independent variable $t$ is the coordinate time TCB. A detailed discussion on the consequences of using TDB instead of TCB is given by \cite{Klioner2008}.

\subsection{Proper time dynamics}
\label{paper4sec:properTimeDynamics}
Eq. (\ref{eq:paper4fundamentalEomIncludingNonGrav}) describes the evolution of the four-velocity $dx^{\mu}/d\tau$ of an observer, with the first equation $(\mu=0)$ describing the evolution of the time component and the other three equations $(\mu=1,2,3)$ describing the translational dynamics. The norm of the four-velocity is constant by definition, \citep[\textit{e.g.}, ][]{MisnerEtAl1973}, {a condition that can be expressed as follows using Eq. (\ref{eq:paper4metricPerturbation}):}
\begin{align}
\frac{d\tau}{dt}=&\sqrt{-g_{\mu\nu}\frac{\dot{x}^{\mu}\dot{x}^{\nu}}{c^{2}}}=\sqrt{1-\frac{v^{2}}{c^{2}}-h_{\mu\nu}\frac{\dot{x}^{\mu}\dot{x}^{\nu}}{c^{2}}}=1+O(c^{-2})\label{eq:paper4properTimeEquationWithMetricPerturbation} 
\end{align}
Since the leading-order difference between proper time rate and coordinate time rate is at the $c^{-2}$ level, we will propagate the proper time difference $\Delta\tau$ {of an observer}:
\begin{equation}
\Delta\tau(t)=\tau(t)-t\label{eq:paper4properTimeDifferenceEq}
\end{equation}
To evaluate the derivative{ of Eq. (\ref{eq:paper4properTimeDifferenceEq}) w.r.t. }$t$ using Eq. (\ref{eq:paper4properTimeEquationWithMetricPerturbation}), we expand it using a Taylor-series, thereby canceling the terms of $O(1)${. This prevents }rounding errors from limiting the precision of our results. The order of this expansion may be chosen such that the error that it introduces is below numerical precision. For a first-order expansion:
\begin{equation}
\frac{d\Delta\tau}{dt}=-\frac{1}{2}\left(\frac{v^{2}}{c^{2}}+h_{\mu\nu}\frac{\dot{x}^{\mu}\dot{x}^{\nu}}{c^{2}}\right)\label{eq:paper4properTimeEquationWithMetricPerturbationFirstOrder}
\end{equation}
By numerically integrating Eqs. (\ref{eq:paper4expandedEom}) and (\ref{eq:paper4properTimeEquationWithMetricPerturbationFirstOrder}), we obtain a numerical solution for the space-time behaviour of a given observer. This is equivalent to the approach taken in {the} generation of the INPOP ephemerides \citep{FiengaEtAl2009}.

\subsection{Time-transfer observable}
\label{paper4sec:timeTransferObservable}
To include time dynamics{ in the estimation, }measurements of the relative proper time rates of separated clocks are required. {Such measurements can} be realized by time transfer. In this section, we summarize the models we use for simulating time-transfer observables, with extensive details provided by \textit{e.g.}, \citep{Klioner1992,PetitWolf1994,BlanchetEtAl2001}. 

We consider two clocks between which time transfer is performed by means of electromagnetic signal transmission,{ for instance laser pulses}. The signal is transmitted by the transmitter (station $A$) at the proper time $\tau^{(A)}(t_{1})$ {and} coordinate time $t_{1}$ and received by the receiver (station $B$) at its proper time $\tau^{(B)}(t_{2})$ {and} coordinate time $t_{2}$. In the following, we shall denote $\tau^{(x)}(t_{i})$ as $\tau^{(x)}_{i}$. The {one-way} time-transfer observable $h^{(1)}_{\tau}$ is then the difference between the two proper times, \citep[\textit{e.g.}, ][]{DuchayneEtAl2009}:
\begin{align}
h^{(1)}_{\tau}=\tau^{(B)}_{2}-\tau^{(A)}_{1}\label{eq:paper4timeTransferObservable}
\end{align}
The proper time rate, given by Eq. (\ref{eq:paper4properTimeEquationWithMetricPerturbation}), will {be different for }the observers $A$ and $B${. This causes} variability of the observable $h_{\tau}^{(1)}$ beyond that{ which is }caused by the translational (and rotational) dynamics of the {receiving and transmitting bodies.}

{The time-transfer observable is calculated by using the numerically integrated translational dynamics and proper times of both the receiver and transmitter from Eqs. (\ref{eq:paper4properTimeEquationWithMetricPerturbationFirstOrder}) and (\ref{eq:paper4expandedEom}):}
\begin{align}
h_{\tau}^{(1)}(\tau^{(A)}_{1})&=\tau^{(B)}_{2}-\tau^{(A)}_{1}\label{eq:paper4TimeTransfer1}\\
t_{1}&=t_{1}(\tau^{(A)}_{1})\label{eq:paper4TimeTransfer2}\\
t_{2}&=t_{1}+T_{21}\left(t_{1};\mathbf{x}_{t}(t),\mathbf{x}_{r}(t)\right)\label{eq:paper4oneWayRangeEq}\\
\tau^{(B)}_{2}&=\tau^{(B)}_{2}(t_{2})\label{eq:paper4TimeTransfer4}
\end{align}
In the above $T_{21}$ denotes the classical one-way light-time observable \citep{Moyer2005}:
\begin{align}
T_{21}(t_{1})&=\frac{\left|\mathbf{x}_{B}(t_{1}+T_{21})-\mathbf{x}_{A}(t_{1})\right|}{c}+\Delta T_{21}\left(t_{1},t_{2},\mathbf{x}_{1},\mathbf{x}_{2}\right)\label{eq:paper4lightTimeEquation}
\end{align}
where the $\Delta T_{21}$ term denotes light time corrections due to atmospheric and relativistic effects. {The model for the one-way range observable that is typically used in orbit determination is calculated from $cT_{21}$, and clearly differs from the one-way time-transfer observable $h_{\tau}^{(1)}$ by virtue of Eqs. (\ref{eq:paper4TimeTransfer2}) and (\ref{eq:paper4TimeTransfer4}).  Note that the time-transfer observable is referenced to a proper time at one of the link ends, whereas as the range observable is referenced to one of the coordinate times \citep{HeesEtAl2012}.}

The observable $h^{(1)}_{\tau}$ may  be related to the one-way light time as:
\begin{align}
h^{(1)}_{\tau}=T_{21}(t_{1}(\tau^{(A)}_{1}))-\int_{t_{0}}^{t_{1}}\frac{d\Delta\tau^{(A)}}{dt}dt+\int_{t_{0}}^{t_{2}}\frac{d\Delta\tau^{(B)}}{dt}dt\label{eq:paper4oneWayWrittenOut}
\end{align}
{where $t_{0}$ denotes the reference time at which the initial conditions for the numerical solution of Eqs. (\ref{eq:paper4expandedEom}) and (\ref{eq:paper4properTimeEquationWithMetricPerturbationFirstOrder}) are specified. From this equation, it is clear that the integrated value of $\Delta \tau$ over long time periods at both the transmitter and receiver influences the observable.} 

When using both an up- and downlink one-way range measurement to construct a two-way observable $h_{\tau}^{(2)}$, for instance using an {(}asynchronous{)} active laser ranging system \citep{Degnan2002,BirnbaumEtAl2010}, the influence of relativistic clock effects is substantially reduced, in the same manner that the influence of clock noise at both link ends is substantially reduced \citep{DirkxEtAl2015}. 
The downlink of a two-way measurement from the space segment to the ground station consists of a signal transmitted by the space segment at coordinate time $t_{3}$, where $t_{3}=t_{2}+\delta t$, with $\delta t$ typically at most on the order of seconds. This signal is then received by the ground station at coordinate time $t_{4}$. The raw two-way light time, as measured by the ground station, then becomes:
\begin{align}
h_{\tau}^{(2)}&=\tau^{(A)}_{4}-\tau^{(A)}_{1}\label{eq:paper4twoWay}\\
&=T_{21}(t_{1}(\tau^{(A)}_1))+T_{43}(t_{3})+\int_{t_{1}}^{t_{4}}\frac{d\Delta\tau^{(A)}}{dt}dt+\delta t\label{eq:paper4twoWayWrittenOut}\\
t_{3}&=t_{1}(\tau^{(A)}_1)+T_{21}(t_{1}(\tau^{(A)}_1))+\delta t
\end{align}
Since the retransmission time at the space segment $\delta t$ is small, the influence of the proper time rate of the space segment on $h_{\tau}^{(2)}$ is almost completely removed. However, the conversion from $h_{\tau}^{(2)}$ to the two-way coordinate light time $t_{4}-t_{1}$  remains sensitive to the difference in rates between $t$ and $\tau^{(A)}$, but only as integrated over $t_{4}-t_{1}$. Comparing Eqs. (\ref{eq:paper4oneWayWrittenOut}) and (\ref{eq:paper4twoWayWrittenOut}), it is clear that the one-way observable is sensitive to clock effects accumulated over long periods of time, whereas the two-way observable is largely insensitive to relativistic clock variations. Nevertheless, since an (asynchronous) two-way observable simply consists of two one-way observables, it can be used to exploit relativistic clock effects during the estimation by including one or both of the one-way time transfer observables separately {into} the estimation.

Since our purpose in this work is to present a model for the space-time dynamics estimation, we do not provide detailed models for relativistic effects on the $\Delta T_{21}$ component of time transfer. Such models have been developed for a variety of contexts and applications, \citep[\textit{e.g.}, ][]{Klioner1992,KopeikinSchafer1999,LinetTeyssandier2002}, but their inclusion does not alter the formulation of our estimation framework. {The inclusion of these effects will slightly change the exact light-time and therefore cause a small modification of $t_{2}$ and $\tau_{2}^{B}$ for a one-way range observable, see Eq. (\ref{eq:paper4oneWayRangeEq}). However, neither the translational nor the time dynamics of the observers is influenced by these effects.}

\section{Coupled space-time estimation}
\label{paper4sec:coupledEstimation}

In this section, we present our proposed{ approach for }coupled space-time dynamics estimation{. We} reiterate that this approach may become important for near-future situations where the same data are used for the estimation of the translational dynamics and the characterization of relativistic clock effects. We start by briefly outlining the problem in Section \ref{paper4sec:coupledIntro}. Subsequently, we derive variational equations for the coupled estimation in Section \ref{paper4sec:relVarEq} and the complete partial derivatives of the time-transfer observables in Section \ref{paper4sec:timeTransferObsPartials}.

\subsection{Proper time estimation}
\label{paper4sec:coupledIntro}
To estimate the orbit of a celestial body or artificial satellite from tracking data, the initial state of that body is adjusted {in such a manner that the (weighted r.m.s.) difference between the measured and modelled observables is minimized} \citep[\textit{e.g.}, ][]{MontenbruckGill2000}. Additionally, parameters that influence the force and/or observation models may be concurrently estimated. 

By extending this approach to the coupled space-time dynamics case, we include the estimation of the initial proper time difference(s) $\Delta\tau$ of one or more observers. The value of $\Delta \tau(t_{0})=\Delta \tau_{0}$ {of these observers }can then be estimated by minimizing the residuals between the modelled and measured values of $\Delta\tau(t)$, as well as any other parameters that influence the solution of Eq. (\ref{eq:paper4properTimeDifferenceEq}), in the same manner as is done in translational orbit determination \citep[\textit{e.g.}, ][]{MontenbruckGill2000}. 

{This approach has the advantage that it removes the issue that relativistic clock signal could be interpreted as translational dynamics, without such an effect being quantified by the estimator. Furthermore, it provides a manner in which the capabilities of relativistic clock signals to enhance the science return from space mission tracking data} \citep{MayrhoferPail2012,Angelil2014,ScharerEtAl2014} {can be robustly analyzed.}  Crucially, the adjustment of the proper time estimation is done concurrently with that of the translational state behaviour, allowing correlations between the two to be identified and {quantified}.

\subsection{{Coupled variational equations}}
\label{paper4sec:relVarEq}

We derive a set of equations to estimate the initial state of $N$ bodies and the proper time differences of $M$ observers. We denote the translational state of a single body $n$ as $\mathbf{y}_{n}=\left(\mathbf{x}_{n};\dot{\mathbf{x}}_{n}\right)$, with $\mathbf{x}_{n}$ the current position of body $n$, and $\dot{\mathbf{x}}_{n}$ its velocity. The concatenated vector of all states is denoted as $\mathbf{Y}$, so that $\mathbf{Y}=\left(\mathbf{y}_{1};...; \mathbf{y}_{N}\right)$. The concatenated vector of all proper time differences is denoted $\mathbf{T}$, so that $\mathbf{T}=\left(\Delta\tau_{1};...; \Delta\tau_{M}\right)$. Finally, we construct our full state vector $\mathbf{X}$ as follows:
\begin{equation}
\mathbf{X}=\begin{pmatrix}\mathbf{Y}\\ \mathbf{T}\end{pmatrix}=\begin{pmatrix}\mathbf{y}_{1}... \mathbf{y}_{N}, \Delta\tau_{1} ... \Delta\tau_{M} \end{pmatrix}^{T}\label{eq:paper4fullStateVector}
\end{equation}
Our goal in this section is to derive equations {to determine} the (linearized) change in $\mathbf{X}$, due to a change in its initial value $\mathbf{X}(t_{0})$ {(}denoted $\mathbf{X}_{0}${)} and the model parameter vector $\mathbf{p}$. This is in contrast to the typical approach to orbit determination and time transfer, where the adjustment of $\mathbf{Y}_{0}$ to the data is performed using an \textit{a priori} fixed $\mathbf{T}_{0}$.

{These changes are }encoded in the state transition matrix $\Phi(t;t_{0})$ and sensitivity matrix $S(t)$ \citep{MontenbruckGill2000}. Extending these matrices to our case, using Eq. (\ref{eq:paper4fullStateVector}) as a state vector, we obtain the following.
\begin{align}
\Phi(t;t_{0})&=\frac{\partial \mathbf{X}(t)}{\partial \mathbf{X}_{0}}=\begin{pmatrix}\frac{\partial \mathbf{Y}(t)}{\partial \mathbf{Y}_{0}} & \frac{\partial \mathbf{Y}(t)}{\partial \mathbf{T}_{0}} \\ \frac{\partial \mathbf{T}(t)}{\partial \mathbf{Y}_{0}} & \frac{\partial \mathbf{T}(t)}{\partial \mathbf{T}_{0}} \end{pmatrix}\\
S(t)&=\frac{\partial \mathbf{X}(t)}{\partial \mathbf{p}}=\begin{pmatrix}\frac{\partial \mathbf{Y}(t)}{\partial \mathbf{p}} \\ \frac{\partial \mathbf{T}(t)}{\partial \mathbf{p}} \end{pmatrix}
\end{align}
These partial derivatives denote the change in the state $\mathbf{X}$ that would occur if {the} initial state or model parameter is changed.

The differential equation{ describing the behaviour of} $\Phi(t,t_{0})$ and $S(t)$ is given by \citep{MontenbruckGill2000}. Here, we generalize their result to the estimation of $N$ {translational} states and $M$ observer proper times. This results in the following formulation for the variational equations:
\begin{align}
\frac{d}{dt}\left(\Phi\left(t;t_{0}\right), S\left(t\right)\right)&=
\begin{pmatrix}\frac{\partial \dot{\mathbf{Y}}(t)}{\partial \mathbf{Y}} & \frac{\partial \dot{\mathbf{Y}}(t)}{\partial \mathbf{T}} \\ \frac{\partial \dot{\mathbf{T}}(t)}{\partial \mathbf{Y}} & \frac{\partial \dot{\mathbf{T}}(t)}{\partial \mathbf{T}} \end{pmatrix}
\left(\Phi\left(t,t_{0}\right), S\left(t\right)\right)+...\nonumber\\
&...+\begin{pmatrix}0_{6N\times(6N+M)} &\frac{\partial\dot{\mathbf{Y}}(t)}{\partial\mathbf{p}} \\
0_{M\times(6N+M)} &\frac{\partial\dot{\mathbf{T}(t)}}{\partial\mathbf{p}}\end{pmatrix}\label{eq:paper4variationalEquations}\\
&=A(\mathbf{X},\dot{\mathbf{X}},t)\left(\Phi\left(t,t_{0}\right), S\left(t\right)\right)+B(\mathbf{X},\dot{\mathbf{X}},t)\\
\Phi(t_{0},t_{0})&=1_{(6N+M)\times (6N+M)}\\
S(t_{0})&=0_{(6N+M)\times N_{P}}
\end{align}
where $N_{P}$ denotes the length of the estimated parameter vector $\mathbf{p}$. 

There is no explicit dependency of $\dot{\mathbf{Y}}$ on any of the {observers'} proper times $\mathbf{T}$, only on the coordinate time $t$. Similarly, the metric in the global coordinates does not depend explicitly on the proper time of a specific observer \citep{SoffelEtAl2003}. {As a result}, there is also no explicit dependency of $\dot{\mathbf{T}}$ on $\mathbf{T}$ and the matrix $A(\mathbf{X},\dot{\mathbf{X}},t)$ reduces to:
\begin{equation}
A(\mathbf{X},\dot{\mathbf{X}},t)=\begin{pmatrix}\frac{\partial \dot{\mathbf{Y}}(t)}{\partial \mathbf{Y}} & 0\\ \frac{\partial \dot{\mathbf{T}}(t)}{\partial \mathbf{Y}} & 0 \end{pmatrix}
\end{equation}

An adjustment in $\mathbf{T}_{0}$ will only result in a change in $\mathbf{T}(t)$, not of $\mathbf{Y}(t)$. However, an adjustment of $\mathbf{Y}_{0}$ of $\mathbf{p}$ influences both $\mathbf{Y}(t)$ and $\mathbf{T}(t)$, introducing the space-time dynamics coupling into the estimation. 

The expression for the top-left matrix block of $A(\mathbf{X},\dot{\mathbf{X}},t)$ 
is composed of the blocks {containing} the derivative {of the state }of body $m$ w.r.t. the state of body $n$ as follows:
\begin{align}
\frac{\partial \dot{\mathbf{Y}}_{m}(t)}{\partial \mathbf{Y}_{n}(t)}=\begin{pmatrix} 0_{3\times 3} & \delta_{mn}1_{3\times 3}\\
\frac{\partial\ddot{\mathbf{x}}_{m}}{\partial \mathbf{x}_{n}} & \frac{\partial\ddot{\mathbf{x}}_{m}}{\partial \dot{\mathbf{x}}_{n}}\end{pmatrix}
\end{align}

To evaluate the components of the variational equations, the partial derivatives of the accelerations and proper time {rates} w.r.t. the full parameter vector $\mathbf{q}(=\left(\mathbf{X}_{0};\mathbf{p}\right))$ must be evaluated. For the equations of motion in the form of Eq. (\ref{eq:paper4condensedEom}), this results in the following general formulation:
\begin{align}
\frac{\partial\ddot{x}^{i}}{\partial\mathbf{q}}&=-\frac{\partial\Gamma^{i}_{\alpha\beta}}{\partial\mathbf{q}}\dot{x}^{\alpha}\dot{x}^{\beta}+
\frac{1}{c}\frac{\partial\Gamma_{\alpha\beta}^{0}}{\partial\mathbf{q}}\dot{x}^{\alpha}\dot{x}^{\beta}\dot{x}^{i}-2\Gamma^{i}_{\alpha\beta}\frac{\partial\dot{x}^{\alpha}}{\partial\mathbf{q}}\dot{x}^{\beta}+...\nonumber\\
&...+\frac{1}{c}\Gamma_{\alpha\beta}^{0}\left(\dot{x}^{\alpha}\dot{x}^{\beta}\frac{\partial\dot{x}^{i}}{\partial\mathbf{q}}+2\frac{\partial\dot{x}^{\alpha}}{\partial\mathbf{q}}\dot{x}^{\beta}\dot{x}^{i}\right)\label{eq:accPartialsChristoffel}
\end{align}
where the symmetry of the Christoffel symbols in the indices $\alpha\beta$ has been exploited. 

Estimation of the initial proper time difference requires the calculation of the partial derivative of the proper time rate $d\tau/dt$ w.r.t. $\mathbf{q}$. These partial derivative are derived from Eq. (\ref{eq:paper4properTimeEquationWithMetricPerturbation}) to be:
\begin{align}
\frac{\partial\dot{\tau}}{\partial \mathbf{q}}&=\frac{1}{2\dot{\tau}}\left(\frac{\partial g_{\mu\nu}}{\partial \mathbf{q}}\dot{x}^{\mu}\dot{x}^{\nu}+2g_{\mu\nu}\dot{x}^{\mu}\frac{\partial \dot{x}^{\mu}}{\partial \mathbf{q}}\right)\label{eq:paper4properTimePartial}
\end{align}

The partial derivatives of the Christoffel symbols, required for the evaluation of Eq. (\ref{eq:accPartialsChristoffel}), may be obtained numerically or analytically
from Eq. (\ref{eq:paper4christoffelMetricLink}) {as follows}:
\begin{align}
\frac{\partial\Gamma_{\alpha\beta}^{\mu}}{\partial \mathbf{q}}=\frac{1}{2}\left(-\frac{\partial h^{\mu\nu}}{\partial\mathbf{q}}\left(h_{\nu\alpha,\beta}+h_{\nu\beta,\alpha}-h_{\alpha\beta,\nu}\right) \right.+...\nonumber\\
...+\left.g^{\mu\nu}\left(\frac{\partial h_{\nu\alpha,\beta}}{\partial\mathbf{q}}+\frac{\partial{h_{\nu\beta,\alpha}}}{\partial\mathbf{q}} -\frac{\partial h_{\alpha\beta,\nu}}{\partial\mathbf{q}}\right)\label{eq:paper4christoffelPartial}
\right)
\end{align}

{From these equations}, the variational equations may be evaluated analytically or numerically from only (first and second) derivatives of the metric (perturbation), in addition to the current state $\mathbf{Y}(t)$ and the metric itself.

\subsection{Time-transfer observable partials}
\label{paper4sec:timeTransferObsPartials}
To incorporate the time-transfer {observable} into the orbit determination, we require the partial derivatives of the $h_{\tau}^{(1)}$ {observable}, given {by} Eq. (\ref{eq:paper4timeTransferObservable}), w.r.t. the full state vector $\mathbf{X}$ and the model parameters $\mathbf{p}$. Denoting (an arbitrary combination of entries of) these vectors by $\mathbf{q}$, the following {is needed}:
\begin{align}
\frac{\partial h_{\tau}^{(1)}}{\partial \mathbf{q}}&=\frac{\partial \tau^{(B)}_{2}}{\partial \mathbf{q}}-\frac{\partial \tau^{(A)}_{1}}{\partial \mathbf{q}}\label{eq:paper4generalTimeTransferPartial}
\end{align}
As discussed in Section \ref{paper4sec:timeTransferObservable}, we reference the time transfer to either the transmission or reception proper time. Here, we will again show the case where we reference the observation to the transmission time, so that $h_{\tau}^{(1)}=h_{\tau}^{(1)}(\tau^{(A)}_{1})$, with a fixed $\tau^{(A)}_{1}${. This makes} the second term on the right-hand side of Eq. (\ref{eq:paper4generalTimeTransferPartial}) equal to zero. The first term on the right-hand side {then becomes:}
\begin{align}
\frac{\partial \tau^{(B)}_{2}}{\partial \mathbf{q}}&=\dot{\tau}^{(B)}_{2}\frac{\partial t_{2}}{\partial\mathbf{q}}+\frac{\partial \tau^{(B)}_{2}}{\partial\mathbf{q}}|_{t_{2}=\text{const}}\\
&=\dot{\tau}^{(B)}_{2}\frac{\partial t_{2}}{\partial\mathbf{q}}+\frac{\partial \Delta\tau^{(B)}_{2}}{\partial\mathbf{q}}
\end{align}
{The first term } on the right-hand side {of these equations }represents the change in reception proper time due to a change in reception coordinate time, and the second term represents the inherent change in the receiver proper time, \textit{i.e.} the direct influence on the integrated result of Eq. (\ref{eq:paper4properTimeEquationWithMetricPerturbationFirstOrder}). The reception coordinate time partial is expanded as follows from Eq. (\ref{eq:paper4oneWayRangeEq}):
\begin{align}
\frac{\partial t_{2}}{\partial\mathbf{q}}&=\frac{\partial t_{1}}{\partial\mathbf{q}}+\frac{\partial T_{21}}{\partial\mathbf{q}}+\frac{\partial T_{21}}{\partial t}\frac{\partial t_{1}}{\partial\mathbf{q}}
\end{align}
The partial derivative of the transmission coordinate time $t_{i}$ is obtained from Eq. (\ref{eq:paper4properTimeDifferenceEq}):
\begin{equation}
\frac{\partial t_{1}}{\partial\mathbf{q}}=-\frac{\partial \Delta\tau^{(A)}_{1}}{\partial\mathbf{q}}
\end{equation}
{since} $\tau^{(A)}_{1}${ is kept fixed}. Combining these results yields:
\begin{align}
\frac{\partial h_{\tau}^{(1)}}{\partial \mathbf{q}}|_{\tau^{(A)}_{1}=\text{const.}}=\frac{\partial \Delta\tau^{(B)}_{2}}{\partial \mathbf{q}}-\frac{\partial \Delta\tau^{(A)}_{1}}{\partial \mathbf{q}}\left(1+\dot{T}_{21}\right)\dot{\tau}^{(B)}_{2}+\frac{\partial T_{21}}{\partial \mathbf{q}}\dot{\tau}^{(B)}_{2}\label{eq:paper4totalPartial}
\end{align}
The final term of this equation, $\frac{\partial T_{21}}{\partial \mathbf{q}}$, is the one-way range partial, as expanded explicitly by \textit{e.g.}, \cite{Moyer2005}. The first two terms on the right-hand side are the (scaled) terms of the extended state transition matrix, as obtained by (numerical) integration of Eq. (\ref{eq:paper4variationalEquations}). It is these terms which our derivation has shown need to be added to the estimation to exploit the coupling between the space and time dynamics. 

As a result of the formulation of Eq. (\ref{eq:paper4totalPartial}) {and the fact }that $\dot{\mathbf{X}}$ is independent of $\mathbf{T}$, the partial derivatives of the observations w.r.t. the initial proper times behave largely as those of range biases. Deviations from unity of the partial derivatives $\frac{\partial h}{\partial \mathbf{T}}$ instead follow from the $(1+\dot{T}_{21})\dot{\tau}_{r}$ term in Eq. (\ref{eq:paper4totalPartial}).
 
Eq. (\ref{eq:paper4totalPartial}) may be used to determine the direct part of the sensitivity {of the observable to a change} in $\mathbf{p}$ and $\mathbf{Y}$. To compute the total sensitivity to $\mathbf{Y}_{0}$ and $\mathbf{p}$, the state transition and sensitivity matrices that are obtained by (numerical) integration of Eq. (\ref{eq:paper4variationalEquations}) are used{ to obtain the following}:
\begin{align}
\frac{\partial h(t)}{\partial \mathbf{X}_{0}}&=\frac{\partial h}{\partial \mathbf{X}}\Phi(t,t_{0}) \label{eq:paper4FullStateSensitivity}\\
\frac{\partial h(t)}{\partial \mathbf{p}}&=\frac{\partial h}{\partial \mathbf{p}}|_{\mathbf{X}=\text{const.}}+\frac{\partial h}{\partial \mathbf{X}}S(t)\label{eq:paper4FullParameterSensitivity}
\end{align}
These partial derivatives are calculated for each of the observations, from which the least squares parameter adjustment can be performed, as discussed in more detail by \textit{e.g.}, \cite{MontenbruckGill2000}.

\section{Test cases: Mars and Mercury lander}
\label{paper4sec:testCases}
We illustrate the methodology we have described in Section \ref{paper4sec:coupledEstimation} with numerical simulations of Mars and Mercury lander missions performing laser ranging and time transfer to Earth. 

{The simulations we present are \emph{not} an detailed analysis and assessment of the performance of the missions that are consideres}, as is performed by \citep[\textit{e.g.}, ][]{TuryshevEtAl2010,HeesEtAl2012}. They are included here to show the potential influence of the relativistic clock effects on the estimation procedure, {for the case of representative planetary lander missions} using laser ranging. We analyze both Mars and Mercury lander missions to assess the influence of the depth of the space segment inside the gravity well of the Sun. Simulations of tracking data to planetary landers have been performed for a wide variety of mission and tracking data types, \textit{e.g.}, \citep{TuryshevEtAl2010,DehantEtAl2011,DirkxEtAl2014}. Also, models and analyses of time transfer at interplanetary distances have been performed by \textit{e.g.}, \citep{Degnan2002,Nelson2011,PanXie2015}. However, none of these simulations{ have included the coupling between }translational dynamics and relativistic clock effects. It is the goal of the present paper to present the influence of this coupling. For an Earth-orbiting mission, though, a sensitivity analysis of this coupling was performed by \citep{DuchayneEtAl2009}, using an approach different from the one derived here. {Their analysis provides a rigorous sensitivity analysis of the time transfer data for the ACES mission. However, the methods they used do not provide a framework in which the contribution of the clock data to the orbit determination and parameter estimation can be quantified.}

In Section \ref{paper4sec:simsettings} we discuss the settings of our simulations. Subsequently, we present a covariance analysis of the performance of the lander missions in Section \ref{paper4sec:covAn}{. There, we consider} both the cases with and without the space-time dynamics coupling. Finally, we{ compare the influence }of proper time estimation with the influence of \textit{a priori} and iterative \textit{a posteriori} clock (re)calibration in Section \ref{paper4sec:posteriorCalib}.

\subsection{Simulation settings}
\label{paper4sec:simsettings}

The settings for our simulations are the following. During one year of operations, we simulate one 30-minute arc of {one-way laser ranging }data per day from four Earth-based International Laser Ranging Service (ILRS) stations \citep{PearlmanEtAl2002} each, {where we simulate a single observation per minute}. From {these} simulated data, we estimate the following parameters using a least-squares approach \citep[\textit{e.g.}, ][]{MontenbruckGill2000}:
\begin{itemize}
\item PPN parameters $\beta$ and $\gamma$. Nominal values equal to 1 in general relativity.
\item Solar $J_{2}$ gravity field coefficient $J_{2\astrosun}$ (quadrupole moment). Nominal value $\approx 2.2\cdot 10^{-7}$ \citep{MecheriEtAl2004}.
\item Mars/Mercury initial translational state $\mathbf{y}_{M,0}$, nominal values taken from DE430 ephemeris \citep{FolknerEtAl2014}. 
\item Initial proper time difference between space segment and each Earth station $i$, denoted $\Delta\tau^{SC}(t_{0})-\Delta\tau^{(E,i)}(t_{0})$.
\item Mars/Mercury-fixed lander position $\mathbf{r}_{L}^{(M)}$ (both  placed equatorially).
\end{itemize}

To decorrelate the PPN parameter $\gamma$ from the other parameters, its influence on the light-time correction $\Delta T_{21}$ is crucial. Here, we include the Sun's first-order effect on light time, as given by \textit{e.g.}, \cite{Moyer2005}.
{Light time-corrections due to other effects are sufficiently small to be neglected in our conceptual analysis} \citep{MinazzoliChauvineau2011}. 

We do not estimate {empirical} clock {noise} parameters, as is done by \cite{DirkxEtAl2015}, as we assume that the systems are equipped with highly accurate clocks, for which the stochastic clock noise is minimal. 
Nevertheless, it should be understood that the detection of long-periodic signals requires highly stable clocks. To be able to observe a clock effect which manifests itself as a 1 cm change in $c\cdot h_{\tau}^{(1)}$ over a period of 1 day requires an Allan deviation of about $4 \cdot10^{-16}$ over such an integration time, which is about a factor of 2 better than that of the proposed deep space atomic clock (DSAC) \citep{PrestageWeaver2007}. For a similar 1-cm change over a period of 1 year, an Allan deviation of the clock of $10^{-18}$ over a similar period is required, which may be facilitated in the future by optical clocks, \citep[\textit{e.g.}, ][]{BloomEtAl2014}. {A detailed analysis of the error sources of interplanetary laser ranging is given by} \cite{Dirkx2015}{}

Summarizing, our simulations are performed as follows:
\begin{itemize}
\item The translational and time dynamics for each observer {are} concurrently numerically integrated using Eqs. (\ref{eq:paper4condensedEom}) and (\ref{eq:paper4properTimeEquationWithMetricPerturbationFirstOrder}), with the full metric given by \cite{SoffelEtAl2003}. We include the point-mass potential of the Earth and Mars/Mercury and the Sun's gravity field including its $J_{2}$ term.
\item From the simulated space-time dynamics, we use Eq. (\ref{eq:paper4timeTransferObservable}) to simulate the observables between the space segment {and the ground stations}.
The observations are constrained by a 15$^{\circ}$ minimum elevation angle and a 5$^{\circ}$ Sun avoidance angle.
\item We numerically integrate the state transition and sensitivity matrices using Eq. (\ref{eq:paper4variationalEquations}). We obtain the partial derivatives of the Christoffel symbols, which are required for evaluating {Eq. }(\ref{eq:paper4christoffelPartial}), by means of numerical differentiation.
\item Using Eq. (\ref{eq:paper4totalPartial}), we calculate the direct contribution of the partial derivatives of the observations w.r.t. the estimated parameters. Subsequently, we use the numerically integrated $\Phi(t,t_{0})$ and $S(t)$ matrices to calculate the complete partial derivative of each observation w.r.t. each of the estimated parameters, using Eqs. (\ref{eq:paper4FullStateSensitivity}) and (\ref{eq:paper4FullParameterSensitivity}).
\item From the partial derivatives of the observations, we set up the information matrix $H$, from which we calculate the covariance matrix \citep{MontenbruckGill2000}. 
\end{itemize}

For the Mercury lander, we evaluate two different mission lengths. Since Mercury's rotation is in a 3:2 resonance with its orbit, a Mercury lander experiences {long} periods of complete darkness and intense sunlight. Therefore, a four-year Mercury lander mission is highly challenging with current technological and budgetary constraints. A 30-day mission, however, would be easier to realize, as it is possible to keep the lander under the same illumination conditions for this amount of time while retaining constant visibility to the Earth. Nevertheless, the depth of Mercury in the Sun's gravity well makes it an attractive test case for exploiting relativistic effects. Since our primary goal in this section is to get a broad overview of the potential effect of space-time dynamics coupling on range data inversion, we will analyze both a 30-day and a 4-year mission, acknowledging the technological difficulties associated with a 4-year mission. 

\subsection{Covariance analysis results}
\label{paper4sec:covAn}
We perform a covariance analysis to investigate the effect of the addition of the $\Delta\tau$ terms in Eq. (\ref{eq:paper4totalPartial}), as well as the addition of the estimation of $\Delta\tau^{SC}(t_{0})-\Delta\tau^{E,i}(t_{0})$ on the formal error and correlation statistics \citep{MontenbruckGill2000}. We present the results of the covariance analyses of the Mars and Mercury lander estimations with the settings defined in Section \ref{paper4sec:simsettings}. {By using a covariance analysis, we implicitly} assume that our truth and estimation models are equal, and that our observations have errors that are identically and independently distributed with a zero-mean Gaussian distribution.{ A more detailed analysis of the error sources in planetary laser ranging, and methods in which to mitigate them, is given by} \cite{Dirkx2015}. 

{The formal error is obtained from the covariance analysis }will be more optimistic than the true error that will be obtained from real data, \citep[\textit{e.g.}, ][]{MartyEtAl2009,KonoplivEtAl2011,DirkxEtAl2014}. However, by comparing the formal error with and without the contribution of the space-time coupling terms, we identify the{ \emph{relative}} sensitivity of the solution {to the inclusion of the coupled space-time dynamics estimation that we propose here}. Furthermore, our covariance analysis will produce the change in the correlations between the estimation of the parameters due to the inclusion of the space-time coupling, {which can be crucial in the interpretation of the estimation results} (Section \ref{paper4sec:Introduction}). 

\begin{figure*}[tbp!]
\centering
\subfigure[]{
\includegraphics[width=0.30\textwidth]{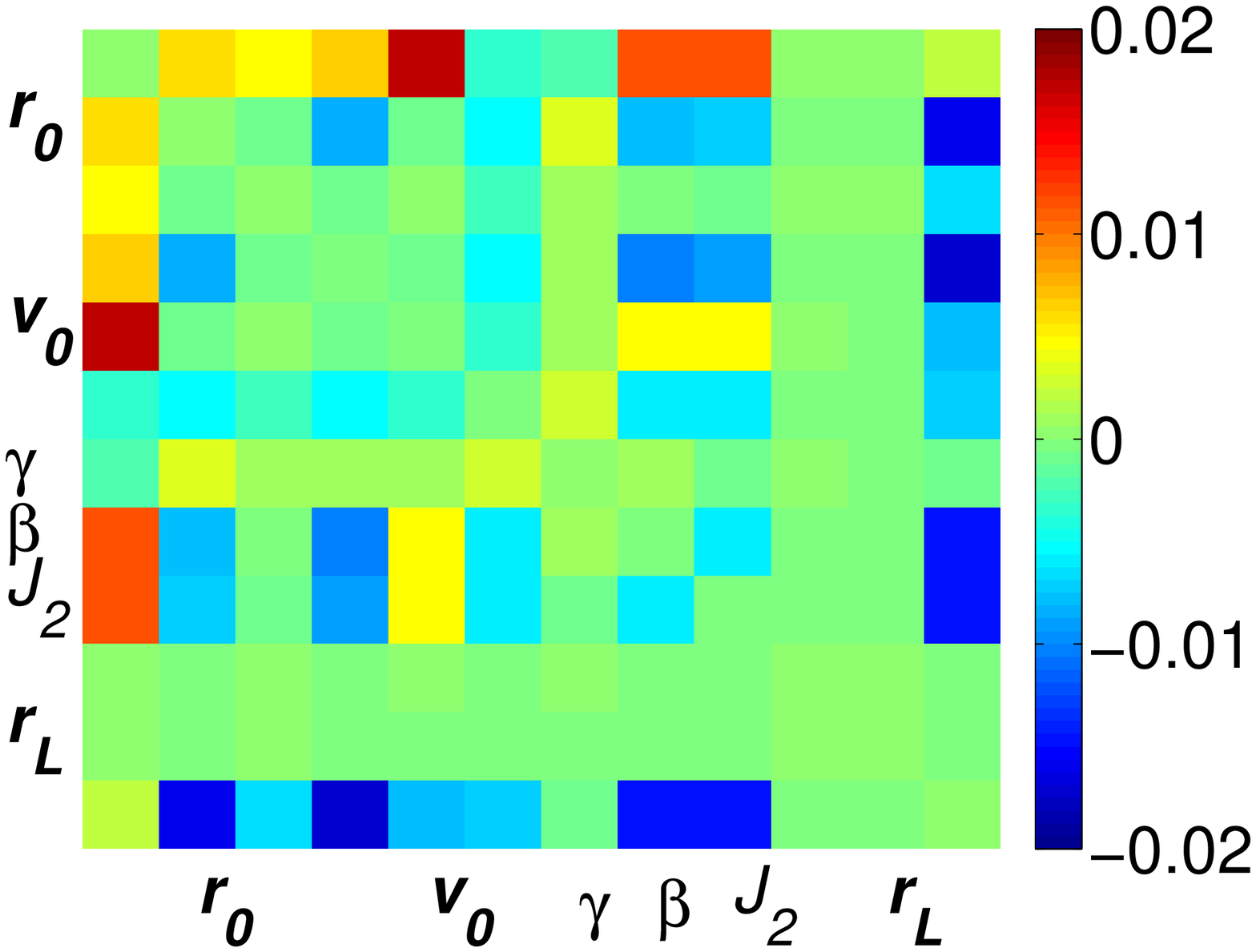}
\label{fig:corrDiffMars}
}
\subfigure[]{
\includegraphics[width=0.30\textwidth]{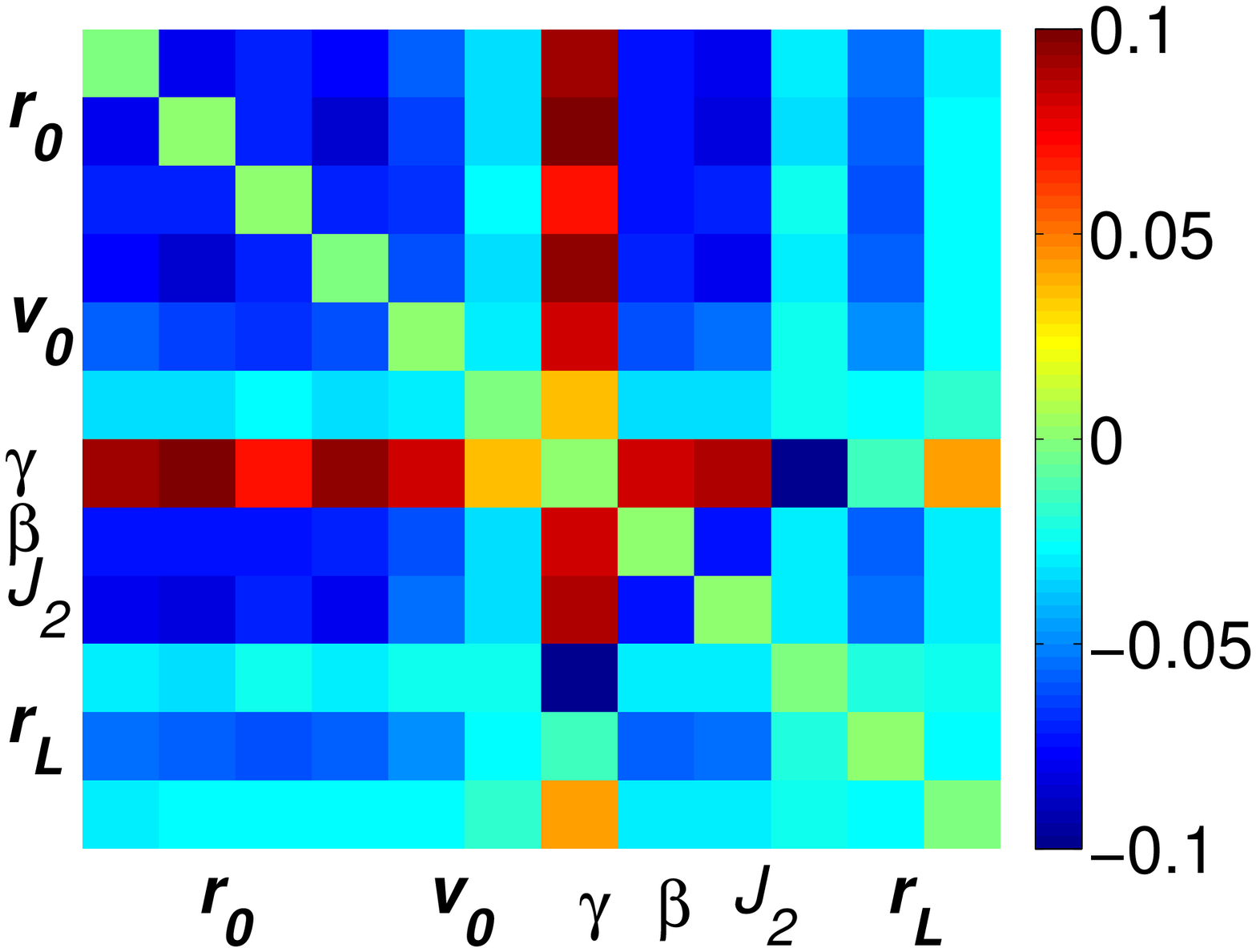}
\label{fig:corrDiffMercury}
}
\subfigure[]{
\includegraphics[width=0.30\textwidth]{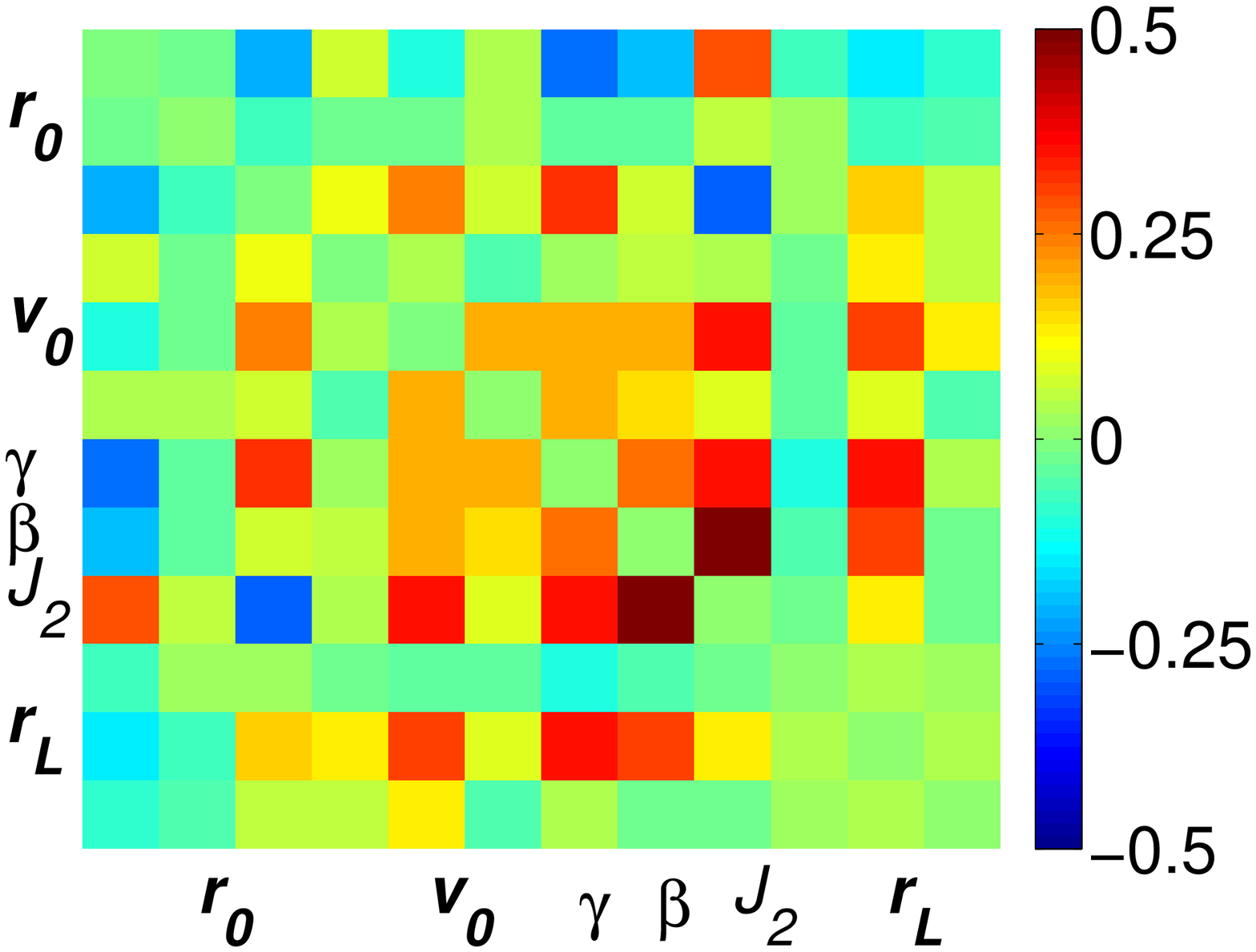}
\label{fig:corrDiffMercuryLong}
}
\caption{Relative difference $\boldsymbol{\Delta}\mathbf{C}$ between correlation matrices (omitting bias entries) of estimation incorporating initial proper time estimation and bias-only estimation, defined in Eq. (\ref{eq:paper4correlationMatrixChange}) a) 4-year Mars mission b) 1-month Mercury mission c) 4-year Mercury mission. }
\label{fig:correlationDiffs}
\end{figure*}

We compare the estimation where we{ include the proper time }estimation with the estimation where we use the classical (no initial proper time estimation) approach, but add the estimation of range biases. As discussed in Section \ref{paper4sec:timeTransferObsPartials}, the behaviour of the observation partials w.r.t. the biases will be similar to that of the initial proper times. The estimation of range biases is typically done in the analysis of satellite laser ranging (SLR) data analysis, although usually over shorter arcs. {In the analyses where we do include }the estimation of the proper time differences, any existing range bias will be absorbed by the proper time difference estimate, so we do not explicitly include the range bias estimation {there}. 

The changes in the entries of the correlation matrix between the time transfer simulations and range simulations will be presented relative to one minus the absolute value of the original (range-only) correlation matrix $\mathbf{C}_{r}$, so that:
\begin{align}
\Delta C_{ij}=\frac{\left|(C_{t})_{ij}\right|-\left|(C_{r})_{ij}\right|}{1-\left|(C_{r})_{ij}\right|}\label{eq:paper4correlationMatrixChange}
\end{align}
where we implicitly set $\Delta C_{ii}$ to 0 and $\mathbf{C}_{t}$ denotes the correlation matrix of the time transfer{ simulation that includes }the proper time difference estimation. In this comparison criterion, we divide by $1-\left|(C_{r})_{ij}\right|$ instead of $\left|(C_{r})_{ij}\right|$, since it is not only important to ascertain the absolute changes in the correlation values, but especially to assess the ability of the estimation to decouple two parameters. That is, a change in correlation of 0.01 is crucial if it changes from 0.99 to 1.0 (or \textit{vice versa}) since it makes the difference between being completely unable and slightly able to decouple the two parameters. Conversely, {a change from} 0 to 0.01 (or \textit{vice versa}){ is} largely irrelevant since the parameters remain almost completely decoupled. 

\begin{table}
\caption{Change in formal errors (in \%) between estimation from range data to landers with estimated initial proper time difference and estimated bias.}
\label{tab:formalErrorDifferences}
\centering
\begin{tabular}{l c c c }
\hline
\hline
& Mars& Mercury & Mercury\\
& (4-years)& (1 month)& (4-years)\\
\hline
$\mathbf{r}_{0}$ & -0.0138&-3.51&2.56\\
$\mathbf{v}_{0}$ & -0.158&-2.31&6.31\\
$\gamma$ & 0.0480&2.93&-9.65\\
$\beta$ &-0.236&-3.68&15.3\\
$J_{2,\astrosun}$ &-0.255&-3.29&9.93\\
$\mathbf{r}_{L}$ & 0.0125&-1.75&3.27\\
\hline
\hline
\end{tabular}
\end{table}

The formal error differences of the estimated parameters with and without the use of the initial proper time difference estimation are shown in Table \ref{tab:formalErrorDifferences}. For the Mars lander, it can clearly be seen that none of the formal errors are significantly affected by the addition of the initial proper time difference estimation, with{ all relative changes} smaller than $1\%$. The{ relative change }in the correlation {matrix} is shown in Fig. \ref{fig:correlationDiffs}.  For the Mars simulation, the difference between the correlation matrices of the two estimations is marginal, as shown in Fig. \ref{fig:corrDiffMars}, with differences around and well below the $1\%$ level. Although these differences are larger than for the formal error, it still signifies the extremely limited influence of the addition of the relativistic clock effects on the estimation.

The difference in formal error statistics for both the short- and long-duration Mercury lander mission is also given in Table \ref{tab:formalErrorDifferences}. Even for the short Mercury mission, the influence of the space-time dynamics coupling on the formal error statistics is more than an order of magnitude larger than for the Mars lander. For the 4-year Mercury mission, the formal error of $\beta$ changes by 15 \%. The formal estimation errors of $\gamma$ and $J_{2,\astrosun}$ both change by about 10 \%. These values show that, also for a long-duration mission relatively deep in the gravity well of the Sun, the differences in formal error will be limited. Nevertheless, the observed differences at the 10 \% level may warrant the use of the combined initial space-time dynamics estimation, but depend on the degree and manner in which non-Gaussian observation uncertainties and model errors propagate into the estimation error budget. 

\begin{figure*}[tbp!]
\centering
\includegraphics[width=0.75\textwidth]{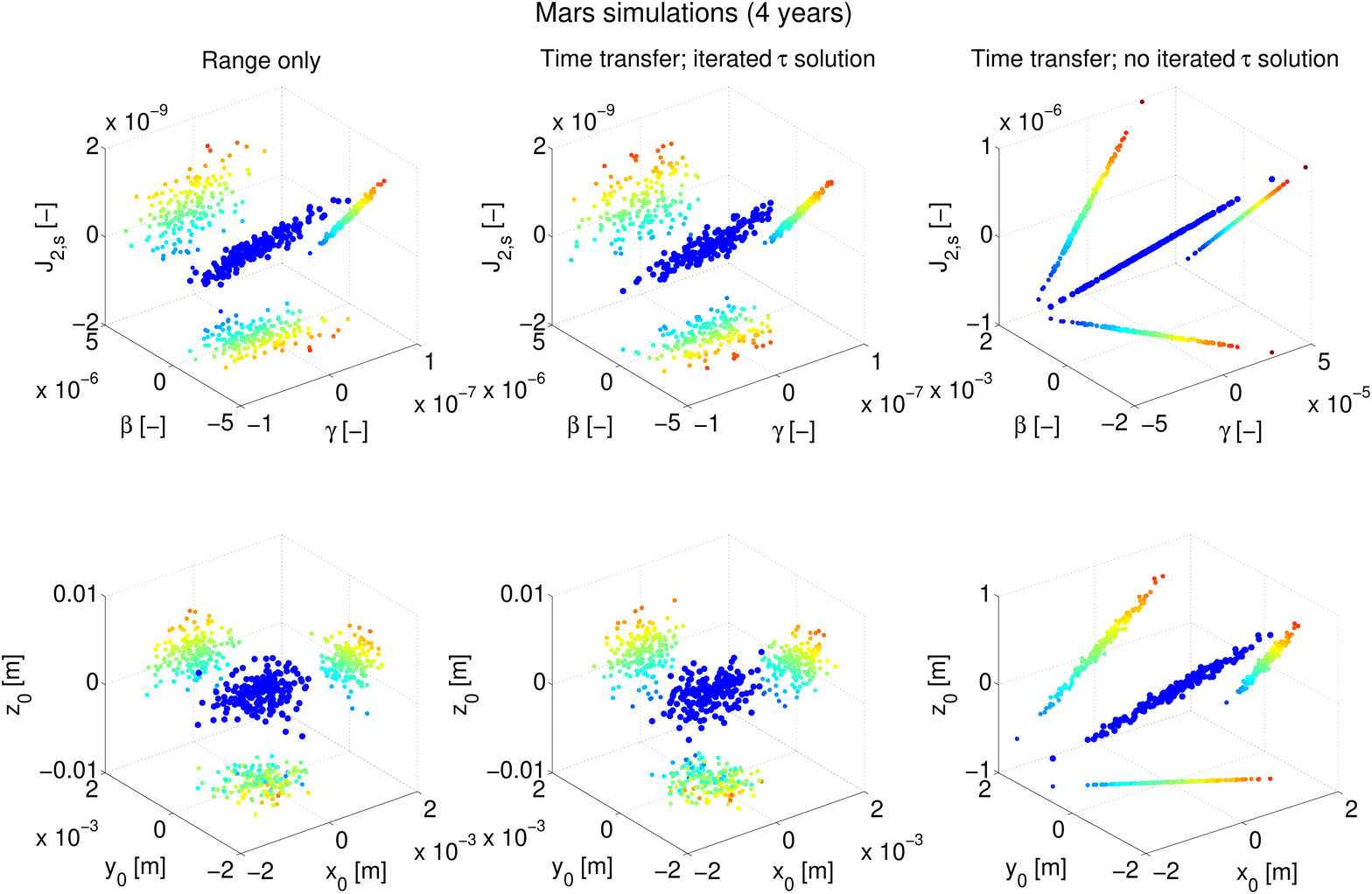}
\caption{Estimation errors obtained from numerical simulations for 4-year Mars mission. Panels show estimation error clouds for ideal $t(\tau)$ conversion, iterated $t(\tau)$ conversion and unupdated \textit{a priori} $t(\tau)$ conversion. Projections of estimations errors onto the three subplanes are also shown, coloured by error in $J_{2,\astrosun}$ (top) and initial $z$-position (bottom).}
\label{fig:marsCalbration}
\end{figure*}

We show the change in the correlation matrix for the Mercury mission, as quantified by $\boldsymbol{\Delta}\mathbf{C}$ from Eq. (\ref{eq:paper4correlationMatrixChange}), in Figs. \ref{fig:corrDiffMercury} and \ref{fig:corrDiffMercuryLong} for the 30-day and 4-year Mercury mission, respectively. As can be seen when comparing these figures to Table \ref{tab:formalErrorDifferences}, the maximum values of $\boldsymbol{\Delta}\mathbf{C}$ are substantially larger than the changes in the formal error (as was the case for the Mars mission simulations), by a factor of about three for both the 30-day and 4-year mission. This difference is due to the $1-(C_{r})_{ij}$ term in Eq. (\ref{eq:paper4correlationMatrixChange}), which causes larger relative {changes in }$\boldsymbol{\Delta}\mathbf{C}$ {for }correlations which are closer to 1 (or -1). For the 4-year mission, relatively large relative changes of up to 0.5 and down to -0.25 are observed{. Positive values} signify cases where the simulations using the coupled estimation have a larger absolute correlation{, see Eq. (\ref{eq:paper4correlationMatrixChange})}. The largest relative change in correlation occurs between $\beta$ and $J_{2,\astrosun}$ for the 4-year mission, where we observe a change in correlation coefficient from $-0.74$ to $-0.87$. This shows that the actual correlation between the signals of these two parameters is significantly larger than that which is obtained from the classical approach (without proper time estimation){. This indicates that}, if the space-time dynamics coupling is omitted in the data analysis process, the correlation between the two parameters that one obtains from the estimation will be substantially more optimistic than the true correlation between the signals from these parameters.

\begin{figure*}[tbp!]
\centering
\includegraphics[width=0.75\textwidth]{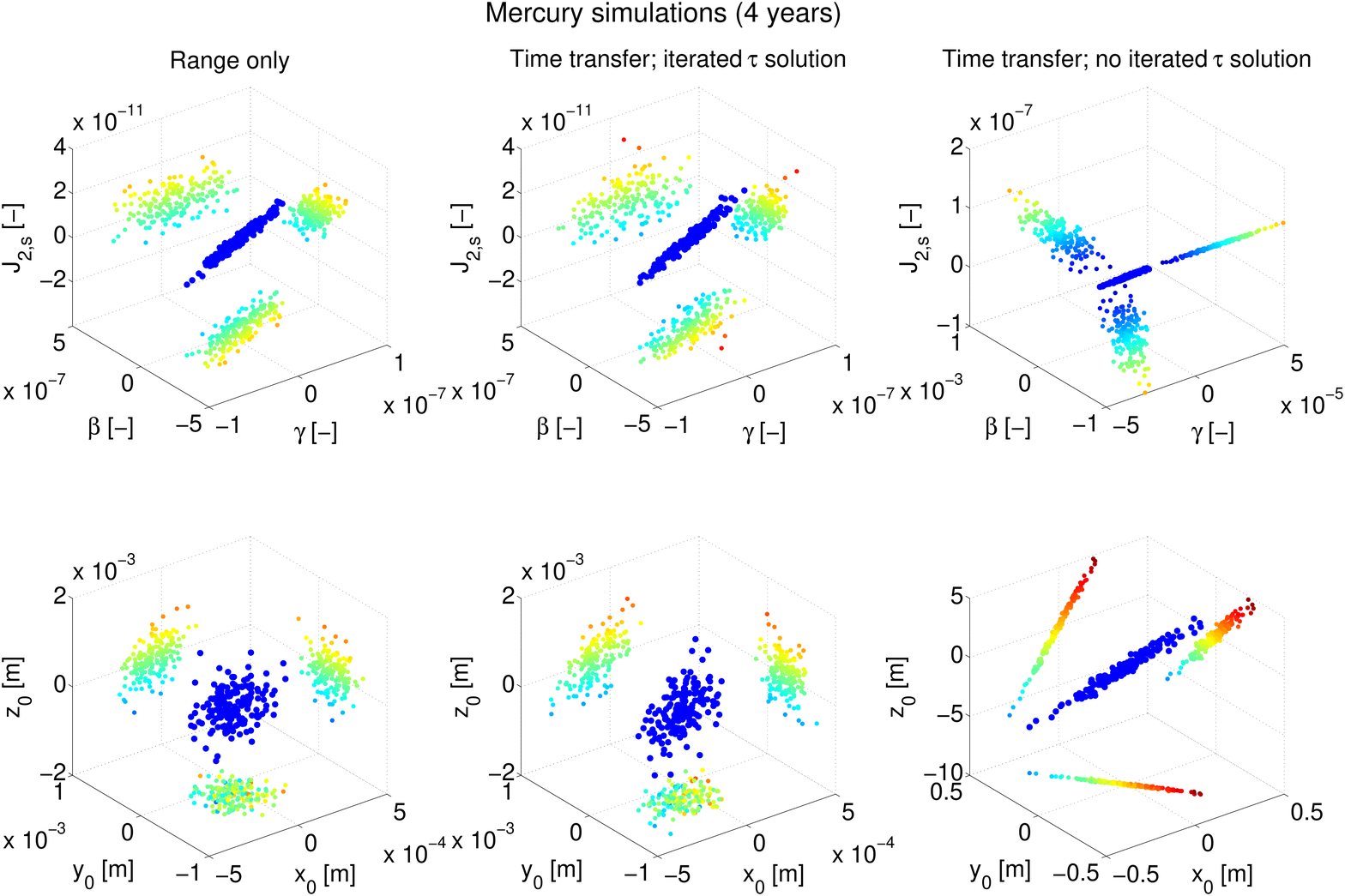}
\caption{Estimation errors obtained from numerical simulations for 4-year Mercury mission. Panels show estimation error clouds for ideal $t(\tau)$ conversion, iterated $t(\tau)$ conversion and unupdated \textit{a priori} $t(\tau)$ conversion. Projections of estimations errors onto the three subplanes are also shown, coloured by error in $J_{2,\astrosun}$ (top) and initial $z$-position (bottom).}
\label{fig:mercuryCalbration}
\end{figure*} 
 
Our results here do not imply that the relativistic clock effects are too weak to be of use for parameter estimation. It instead shows that when using the same data (laser range in these simulations) for orbit determination and time transfer, the resulting orbital accuracy will put stronger constraints on the parameters of interest than the relativistic clock effects. However, this may not be the case for missions where time transfer is only sporadically performed, and orbit determination is also performed using other data types such as Doppler tracking. In these cases, the time-transfer observable may be relatively more accurate than the orbital accuracy, potentially increasing the relative strength of the time-transfer observables in the estimation. Also, highly accurate clocks carried by orbiting spacecraft can be used for long-arc comparison of ground- and space-based clocks, whereas their orbits are typically estimated over relatively short arcs (hours to days) to prevent the accumulation of (non-conservative) force model errors. As such, the longer duration over which the proper time behaviour could potentially be estimated (depending on the clock's stability), compared to the duration over which the state behaviour can be estimated, may increase the importance of the inclusion of the proper time into the estimation.

\subsection{Influence of \textit{a priori} and {a posteriori} calibration}
\label{paper4sec:posteriorCalib}
Having analyzed the influence of the space-time dynamics coupling on the estimation's formal errors in Section \ref{paper4sec:covAn}, we now investigate by direct numerical simulation the influence of omitting this coupling and performing \textit{a priori} or iterative \textit{a posteriori} clock calibration. 

To do so, we perform an estimation using the simulation settings defined in Section \ref{paper4sec:simsettings}.
We perturb the parameters at their current level of uncertainty and use an iterative least-squares method to estimate these parameters. We perform three sets of simulations for both the Mars and Mercury mission:

\begin{enumerate}
\item Estimation in which the true $t(\tau)$ conversion for each observer is known perfectly.
\item Estimation in which a $t(\tau)$ conversion is recomputed following each iteration of the estimation (\textit{a posteriori} clock calibration).
\item Estimation in which a $t(\tau)$ conversion based upon the \textit{a priori} {perturbed} parameter estimate is used.
\end{enumerate}
From these simulations, we will be able to determine the parameter estimation error that is made due to mismodelled relativistic clock effects, when using the typical orbit determination method. 

We find no difference in our results for the 1-month Mercury lander mission between the three cases, due to the fact that the estimation errors of even the ideal case are larger than the initial perturbation in the parameter set. That is, for the short Mercury mission, the tracking data inversion does not result in an improved parameter estimation compared to the \textit{a priori} uncertainty.

For both the Mars and Mercury 4-year tracking simulations, the resulting uncertainties for 100 simulations are shown in Figs. \ref{fig:marsCalbration} and \ref{fig:mercuryCalbration}, respectively. The error clouds of both the initial position of Mars/Mercury and the parameters $\gamma$, $\beta$ and $J_{2,\astrosun}$ are shown. In these figures, the projections of these clouds onto the three planes are also shown. We note that our results for $\gamma$ for the 4-year Mars simulation are similar to the estimation results obtained by \cite{TuryshevEtAl2010} (error of $1.4\cdot 10^{-7}$ after 3 years; $7.9\cdot 10^{-8}$ after 6 years), who perform simulations of laser ranging to Phobos. Our results for $\beta$ and $J_{2}$ are about an order of magnitude more optimistic, most likely due to our smaller set of estimated parameters and resultantly reduced correlations. However, {we reiterate that }our goal here is not {so much} to provide highly accurate absolute estimates for attainable estimation accuracy, but to compare different data analysis techniques.

We observe{ differences of orders of magnitude }between cases 2 and 3. That is, the errors in the \textit{a priori} $t(\tau)$ conversion are sufficiently large to drive the estimation to erroneous results. This influence was not observed in the previous section, where the truth and estimation model were assumed to be equal. This indicates that the use of \textit{a priori} time ephemerides, for instance those of \citep{IrwinFukushima1999,FiengaEtAl2009}, for the data analysis of planetary laser ranging missions can result in substantial degradations of the estimation quality. In general, this will be true if the \textit{a priori} orbital and parameter uncertainty is much greater than the range measurement uncertainty. 

It must be noted, however, that the simulations shown here are performed using purely Gaussian noise on the observations. As a result, the estimation accuracy shown here for cases 1 and 2 is substantially better than what will be obtainable from real tracking data, which will include time-correlated and non-Gaussian noise. Differences between true and formal errors for interplanetary (radiometric) tracking data inversion are typically around a factor 5-10  \citep[\textit{e.g.}, ][]{MartyEtAl2009,KonoplivEtAl2011}.

The difference between cases 1 and 2 is insignificant for the 4-year Mars mission. This is consistent with the fact that we found formal error differences at the 0.1 \% level in Section \ref{paper4sec:covAn}. For the 4-year Mercury mission, we find small but statistically significant differences between the error distributions of cases 1 and 2, with around 10 $\%$ degradation in estimation error for case 2, compared to case 1. Again, this is consistent with the results presented in the previous section, which showed differences in formal errors of around $5-10$ \% for simulations with and without the inclusion of the coupling.

\section{Conclusions}
\label{paper4sec:conclusions}
We have derived an estimation framework with which the influence of relativistic clock effects on (interplanetary) range and time-transfer data analysis can be included on an equal footing with relativistic influence on translational dynamics{. The main }theoretical results of our development being Eqs. (\ref{eq:paper4variationalEquations}) and (\ref{eq:paper4totalPartial}), which are extended versions of the classical equations used in orbit determination. {Our approach allows the degree to which relativistic clock effects can be used to improve parameter estimation and orbit determination to be directly quantified. Also, it provides a reliable quantification of the correlation between the influence of relativistic clock effects and translational dynamics on the time transfer observable, allowing for robust tracking data analysis for a broad range of mission profiles.}

We have simulated{ one-way }data to a Mars and Mercury lander and used the simulated data for a covariance analysis, assessing the possibilities to use these data for the determination of $\gamma$, $\beta$ and $J_{2\astrosun}$. We analyze the use of both range measurements (including the estimation of range biases) and time transfer (including the estimation of an initial proper time difference). The inclusion of the estimation of the proper time difference between Earth stations and a Mars lander has a negligible influence on both the correlations between the estimated parameters and their formal errors, with differences in the formal errors of around 0.1 \% observed. For a Mercury lander this formal error difference rises to 1\% for a 1-month mission and 10 \% for a 4-year mission. However, we find a significantly greater relevant degree of change in the correlation matrix (as quantified by $\boldsymbol{\Delta}\mathbf{C}$) than for the formal errors, which can be of importance for the interpretation of the robustness of the results, as well as any statistical analysis of postfit residuals.

These results indicate that the influence of our estimated parameter set on the translational dynamics is (much) stronger than on the proper time rate,{ due to the $c^{-2}$ term in Eq. (\ref {eq:paper4properTimeEquationWithMetricPerturbationFirstOrder}).}
For spacecraft state estimation, where the translational state is typically determined over short arcs, a highly stable clock may allow substantially modified parameter estimation quality. 

We have shown by direct numerical simulation that using an \textit{a priori} time ephemeris with sufficiently low accuracy can result in orders of magnitude degradation of the true error distribution compared to the formal errors. Recomputing the proper-to-coordinate time conversion for each observer after each iteration of the least-squares estimation results in a degradation similar to the changes {that are observed in the formal errors when including the proper time estimation.}

\begin{acknowledgements}
Dominic Dirkx is financed by the FP7 ESPaCE project, financially supported by the EC FP7 Grant Agreement 263466. The authors thank Alexander Polnarev of the Queen Mary University of London, Sergei Kopeikin of the University of Missouri and an anonymous reviewer for their insightful input, which significantly improved the clarity and quality of the manuscript. 
\end{acknowledgements}
\renewcommand*\thesection{\arabic{section}}